\documentclass[twocolumn,runningheads,final]{svjour2}
\smartqed  
\usepackage{graphicx}
\usepackage{mathptmx}   
\usepackage{color}
\usepackage[authoryear]{natbib}
\journalname{Astrophysics and Space Science (CoRoT/ESTA Volume)}
\newcommand{\alpht  }{\alpha_\mathrm{t}}
\newcommand{\alphs  }{\alpha_\mathrm{s}}

\newcommand{\alphMLT}{\alpha_\mathrm{MLT}}

\newcommand{\Teff   }{T_\mathrm{eff}}

\newcommand{\Hp     }{H_{\!{P}}}
\newcommand{\Cp     }{C_{\!{P}}}

\newcommand{\jt     }{j_\mathrm{t}}
\newcommand{\jw     }{j_{\wturb}}

\newcommand{\delM   }{\Delta m}
\newcommand{\betr   }{\beta_{r}}

\newcommand{\nablmu }{\nabla_{\!{\mu}}}
\newcommand{\nablad }{\nabla_{\!\mathrm{ad}}}
\newcommand{\nablrad}{\nabla_{\!\mathrm{rad}}}
\newcommand{\Pgas   }{P_\mathrm{gas}}

\newcommand{\wturb  }{\bar{\omega}}

\newcommand{\komma  }{\,\,\mbox{,}}
\newcommand{\lderivt}[1]{\frac{\mbox{D} #1}{\mbox{D}t}}
\newcommand{\gradm  }[1]{\frac{\partial #1}{\partial m}}

\newcommand{\gradn  }[1]{\frac{\partial #1}{\partial n}}
\newcommand{\p}[2]{\frac{\partial #1}{\partial #2}}

\newcommand{\od}[2]{\frac{d #1}{d #2}}

\def\myfigure#1#2{
  \begin{figure}
  \begin{center}
  \includegraphics[width=\columnwidth]{#1}
  \end{center}
  \caption{#2}
  \end{figure}
}
\newcommand{\cws}[1]{{#1}}

\begin{document}

\title{YREC: The Yale Rotating Stellar Evolution Code
}
\subtitle{Non-rotating version, seismology applications}


\author{P.~Demarque, D.~B.~Guenther, L.~H.~Li, A.~Mazumdar and
        C.~W.~Straka}


\institute{P.~Demarque, L.~H.~Li, A.~Mazumdar \at 
           Department of Astronomy, Yale University,\\
	   New Haven, CT 06520-8101, USA\\
           \email{pierre.demarque@yale.edu}\\
           \email{li@astro.yale.edu}\\
           \email{anwesh.mazumdar@yale.edu}
	   \and
	   D.~B.~Guenther \at
	   Department of Astronomy and Physics,\\
	   Institute for Computational Astrophysics,\\
	   Saint Mary's University,\\
	   Halifax, Nova Scotia, Canada B3H 3C3\\
	   \email{guenther@ap.smu.ca}
           \and
           C.~W.~Straka \at
           Centro de Astrof{\'i}sica da Universidade do Porto (CAUP),\\
	   Rua das Estrelas,
	   4150-762 Porto, Portugal\\
	   \email{straka@astro.up.pt}
}

\date{Received: date / Accepted: date}

\maketitle
\begin{abstract}
The stellar evolution code YREC is outlined with emphasis on its applications 
to helio- and asteroseismology. The procedure for calculating calibrated solar 
and stellar models is described. Other features of the code such 
as a non-local  
treatment of convective core overshoot, and the implementation of a 
\cws{parametrized}
description of turbulence in stellar models, are considered in some 
detail.  The code has 
been extensively used for  
other astrophysical applications, some of which are briefly mentioned 
at the end of the paper.

\end{abstract}

\section{Introduction}\label{intro}
The aim of this paper is to provide an overview of the Yale Rotating Stellar 
Evolution Code (YREC), as it has been 
applied in the last few years to research in helio- and asteroseismology.  
Although YREC contains extensions to model the effects of rotation 
in an oblate coordinate system, we describe here the ``non-rotating'' version.

In addition to a general description, we shall  
emphasize three features of the code which have been implemented  
because of their 
special relevance to seismology.  The first feature 
is the procedure utilized for the automatic calculation of calibrated 
solar and stellar models whose pulsational properties are to be 
investigated.  The second feature is the treatment of convective
core overshoot. Finally, the third feature is the 
implementation in stellar models of
the effects of turbulence on the structure of the surface layers of 
stars with a convective envelope.  The \cws{parametrization} of turbulence to one 
dimension is based on three-dimensional radiative 
hydrodynamical (3D HRD) simulations of the highly superadiabatic layer 
(SAL) in the atmosphere.  The interaction of turbulent convection 
and radiation in these thin transition 
regions is poorly known.  Oscillation frequencies are sensitively 
affected by the structure of transition regions between 
radiative and convective layers.  Seismology thus offers 
a unique opportunity to explore a long standing problem in stellar physics.  

Like most stellar evolution codes, YREC is a
continuously evolving research tool to which many have contributed. 
As a result, different versions of YREC are in use
at several institutions, which have been applied to a variety of research
purposes. Some of the most significant applications of YREC 
are listed in the text and at the end 
of this paper (see Sect.~\ref{other}).
The rotating version of YREC, originally developed by 
\citet{1988PhDT........14P}, 
includes a 1.5D treatment of rotation, extending the work of 
\citet{1970stro.coll...20K} and \citet{1981ApJ...243..625E}, and using the 
formalism of \citet{1980PhDT.........5L}. 
A 2D version of YREC has also recently 
been implemented, specifically to address some fundamental aspects of  
solar magnetic activity \citep{2006ApJS..164..215L}.

Sect.~\ref{num} outlines the numerical scheme adopted to solve the classical  
differential
equations of stellar structure and evolution. 
The treatment of the boundary conditions, of special importance for 
seismology, are described in Sect.~\ref{bc}.  The
constitutive physics, i.e. the equation of state and radiative and 
conductive opacities,
are reviewed in Sect.~\ref{eos}, and the nuclear processes are described 
in Sect.~\ref{nuc}.  Stellar
physics topics such as superadiabatic convection, element diffusion,  
convective core overshoot, and turbulence in the outer layers, all 
of which also have important
seismological signatures, are covered in 
Sect.~\ref{sbc}, Sect.~\ref{diff}, Sect.~\ref{ov} and 
Sect.~\ref{turb}, respectively.  The operation of the code is described in 
Sect.~\ref{runyrec}, with
emphasis on helio- and asteroseismic applications. 
Seismic diagnostics applications are described in Sect.~\ref{dia}.
The role played by 
YREC in the research on solar neutrinos and helio-seismology 
is summarized in Sect.~\ref{solar}.  
Studies of advanced evolutionary phases and 
applications to stellar population studies are 
listed in Sect.~\ref{other}.

\section{Henyey code}\label{num}
The four first-order simultaneous equations of stellar structure are 
well-known, and have been frequently 
discussed in the literature \citep{1958ses..book.....S}.
YREC uses mass as the independent variable in the formulation of the equations 
(Lagrangian formulation).
The problem is a two-point boundary value problem,
with boundary conditions at
the center and at the surface of the model.  A relaxation
technique, based on a finite difference approximation, is used. 
The method, first applied to the stellar structure problem by 
\citet{1959ApJ...129..628H}, is 
known as the Henyey method.  Useful descriptions 
of the Henyey method are given   
in the paper by \citet{1964ApJ...140..524L} and 
the book by \citet{1990sse..book.....K}.  Specific details about  
the numerical procedures in the YREC implementation   
can be found in the Appendices of \citet{1976PhDT.........8P}, which describe an 
earlier Henyey code on which YREC is based.  \citet{1976PhDT.........8P} 
also provides information about the treatment of the constitutive 
physics, although most of the physics details have  
been updated since then. 
  
In the Henyey method, the model star is divided  
into $n$ concentric shells by means of $n+1$ suitably chosen values of the 
independent variable (mass), or points, in the interval 
defined by the innermost point (near the center) and the outermost point, 
which is specified by the user and 
located at the base of the envelope integrations.  The four 
differential equations are replaced in each shell by approximating difference 
equations 
relating the values of the dependent variables at adjacent points.  
There are four 
dependent variables in each of the $n$ shells, providing a set of $4(n+1)$ 
linear equations which, together with two boundary conditions at the center 
and two at the surface, can 
be solved to determine approximate corrections to the $4(n+1)$ dependent 
variables, starting from a first approximation model.      
The set of simultaneous equation is solved by iteration until the corrections 
in each variable satisfy 
a specified convergence limit.     

\subsection{Shell redistribution}\label{shell}
The shell\cws{s} are distributed so as to optimize numerical accuracy and efficiency.
In order to 
follow the evolution from the earliest gravitational contraction phase all 
the way to the hydrogen and helium shell burning phases, it is necessary to 
redistribute the shells in the model.  This is especially 
critical during shell burning. After a star exhausts its core supply 
of hydrogen it begins burning hydrogen in a shell. Initially the shell 
is almost 
$0.2 M_{\odot}$ thick for a $1 M_{\odot}$ 
star but the shell quickly narrows to only 
$0.001 M_{\odot}$ 
as the star evolves up the giant branch, thinning to 
$0.0001 M_{\odot}$ 
at the point of helium flash. Because of the high temperature 
dependence of helium burning, helium burning shells are even thinner 
than hydrogen burning shells. YREC will add or remove shells according 
to the size of gradients in structure (i.e., pressure, temperature, and 
composition) and gradients in luminosity, as well as the size of 
Henyey corrections applied during the iteration procedure. The code keeps 
track of physically real discontinuities so that they are not smoothed 
during the rezoning process. Interpolation is linear. Our own testing 
has shown that using higher order methods such as oscillatory spline 
interpolation introduces numerical oscillations near the tip of the 
giant branch.

\subsection{Time steps}\label{tsteps}
The models are advanced in time through two terms in the energy equation, 
the nuclear energy term (Sect.~\ref{nuc}), and the time rate of change of entropy due 
to contraction or expansion during evolution. Special care is taken to preserve 
numerical accuracy for small time steps (\citet{1976PhDT.........8P}).

One can either specify the time step or have YREC automatically 
determine the optimum time step during evolution. When producing 
accurately calibrated solar models, to maintain numerical consistency 
it helps to specify the time step interval. In most other situations 
it is best to let YREC determine the time step based on user 
specified convergence tolerance criteria. During nuclear burning 
phases of evolution YREC will guess the time step based on the rate 
at which hydrogen and helium (if applicable), are being consumed in 
each shell of the model. During gravitational contraction phases of 
evolution YREC will control the time steps by monitoring the change 
in temperature, pressure, and luminosity from one model to the next. 
During helium flash if the model fails to converge during a evolutionary 
time step, YREC is also able reduce the time step by a user specified 
factor and redo the evolutionary step. 
More details regarding the operation of YREC can be found in 
Sect.~\ref{runyrec}. 

\section{Boundary conditions}\label{bc}
\subsection{Center}\label{cbc}
The two inner boundary conditions constrain the values of the 
radial distance and luminosity variables at the innermost mass shell.
Because of the false singularity at the center, the innermost point 
is not at the very 
center, but in a shell chosen 
close to the center.  
Note that in order to preserve accuracy, special care must be taken 
with the position of the innermost shell, especially  
in pulsation calculations (see Sect.~\ref{pulse})\cws{.}
\subsection{Surface}\label{sbc}
The outer boundary conditions depend on the structure near the surface.  
Because a model of the outer layers depends on the global properties of the 
star, i.e. its surface gravity $\log~g$ and effective temperature 
$T_\mathrm{eff}$, 
the problem is implicit.  In order to specify 
the surface boundary conditions, which 
relate the variations of the pressure and temperature  
variables to the total luminosity and radius of the star, 
three inward envelope integrations are constructed.  
These envelope integrations are 
chosen so as to form a triangle in the theoretical HR-diagram 
(i.e. the $\log~L/L_{\odot}$ \cws{vs. $\log~\Teff$} plane).  

The inward envelope integrations consist of two main parts.  
The outermost layers,
starting at optical depth near $\tau = 10^{-10}$, which are 
effectively isothermal at 
the start, are described by a 
\cws{gray} radiative atmosphere specified by a $T(\tau)$ relation and integrated 
to the appropriate value of $\tau$ at which the temperature 
reaches \cws{$\Teff$} (e.g. $\tau = 2/3$ for the Eddington approximation, 
$\tau = 0.312156330$ for the \cws{\citet{1966ApJ...145..174}}
atmosphere). This surface  
in the star is usually defined as the \emph{photosphere}.

\cws{As an alternative to the atmosphere integrations, more complex
atmospheres from pre-computed libraries can be also used, such as those
from \citet{1998kurucz}.}

Below the photosphere, all variables but the luminosity variable (which is 
held to be constant in the outer envelope) are 
integrated to a chosen value of the mass. 
The integration is carried out using \cws{$\log~P$} as 
the independent variable, to the value of the mass variable 
at which the surface boundary conditions for the interior are 
computed (the base of the envelope).
The region which extends from this value of the mass to the innermost 
shell of the star constitutes the \emph{interior} of the stellar model.  
In convectively unstable layers of the envelope (as determined 
by the local Schwarzschild 
criterion), the temperature gradient is evaluated 
according to the formalism of \citet{1995ApJ...440..297S}, 
which is designed to describe 
superadiabatic convection.  It is in this region that 
the peak of the highly superadiabatic transition layer (SAL) is 
normally located (as it is in the Sun).  
The main advantage of the \citet{1995ApJ...440..297S} formalism is that 
by a suitable choice of parameters, it can be made 
to reproduce either the standard \emph{mixing length theory} (MLT) 
\citep{1958ZA.....46..108B} 
or the theory of \citet{1992ApJ...389..724C}, sometimes called FST.  
In order to preserve continuity in the convective 
temperature gradient at the envelope-interior 
interface, the \citet{1995ApJ...440..297S} formalism is used to calculate the 
convective gradient both in the envelope and in the interior whenever 
superadiabaticity exceeds a preset value.

Another feature of the envelope integration, 
described in more detail in Sect.~\ref{turb},
includes a 1D \cws{parametrization} of the effects of turbulent pressure and 
turbulent kinetic energy in the outer layers.          
 
\section{Equation of state and opacities}\label{eos}
YREC has been updated regularly so as to incorporate the latest research
developments regarding equation of state and opacity in the
stellar interior, while maintaining backward compatibility with earlier
versions of the same.  The current version uses the latest OPAL
opacities \citep{1996ApJ...464..943I} and OPAL equation of 
state \citep{2002ApJ...576.1064R}.  At
low temperatures ($\log~T < 4.1$) opacities are obtained from the tables
provided by \citet{2005ApJ...623..585F}.

At each mass shell the EOS is obtained by interpolation from the
standard tables.  Since the EOS is weakly dependent on $Z$, we use only
one set of tables at a fixed $Z$, obtained by the $Z$-interpolation
routine provided with the OPAL EOS package. For models with metal
diffusion, the value of $Z$ at which the EOS is interpolated is chosen
at a suitable intermediate value. The EOS quantities at the desired $X$,
$T$ and $\rho$ are obtained by quadratic interpolation from the tables.
The results and the derivatives are smoothed by mixing overlapping
quadratics.  For opacity, a four-point Lagrangian interpolation scheme
is used over a 4-dimensional grid of $Z$, $X$, $T$, and $\rho$.

\section{Diffusion}\label{diff}
The diffusion of chemical elements by gravitational settling
\cws{and thermal diffusion} is implemented following the 
prescription of \citet{1994ApJ...421..828T}. Options in the code include no diffusion,
helium diffusion only, or both \cws{$Y$ and $Z$} diffusion.
The analytical fits 
provided by \citet{1994ApJ...421..828T} can also be used instead of the tabulated 
diffusion coefficients, to speed up the computations.

\section{Nuclear Reactions}\label{nuc}
The nuclear reaction rates in conjunction with the corresponding energy
release ($Q$-values) are important for the evolution of chemical
species, the energy input from nuclear fusion reactions and
for the neutrino fluxes.
The reactions explicitly calculated in YREC are the following:
\begin{eqnarray}
 {{}^1}\mathrm{H}  + {{}^1}\mathrm{H}  & \rightarrow &  {{}^2}\mathrm{H}  + e^{+} + \nu \label{eq:pp}\\
 {{}^3}\mathrm{He} + {{}^3}\mathrm{He} & \rightarrow &  {{}^4}\mathrm{He} + 2\,{{}^1}\mathrm{H} \\
 {{}^3}\mathrm{He} + {{}^4}\mathrm{He} & \rightarrow &  {{}^7}\mathrm{Be} + \gamma
\label{eq:he3he4} \\
 {{}^{7}}\mathrm{Be} + e^{-} + {{}^1}\mathrm{H} & \rightarrow & 2\, {{}^4}\mathrm{He} + \nu
\label{eq:be7e}\\
 {{}^{7}}\mathrm{Be} + {{}^1}\mathrm{H} & \rightarrow & 2\, {{}^4}\mathrm{He} + \gamma + e^{+} + \nu
\label{eq:be7p} \\
 {{}^{12}}\mathrm{C} + {{}^1}\mathrm{H} & \rightarrow &  {{}^{13}}\mathrm{C} + \gamma + e^{+} + \nu
\label{eq:cno1}\\
 {{}^{13}}\mathrm{C} + {{}^1}\mathrm{H} & \rightarrow &  {{}^{14}}\mathrm{N} + \gamma \\
 {{}^{14}}\mathrm{N} + {{}^1}\mathrm{H} & \rightarrow &  {{}^{15}}\mathrm{N} + \gamma + e^{+} + \nu \\
 {{}^{15}}\mathrm{N} + {{}^1}\mathrm{H} & \rightarrow &  {{}^{12}}\mathrm{C} + {{}^4}\mathrm{He}
\label{eq:cno1l}\\
 {{}^{16}}\mathrm{O} + 2\, {{}^1}\mathrm{H} & \rightarrow &  {{}^{14}}\mathrm{N} + {{}^4}\mathrm{He} + \gamma +  e^{+} + \nu
\label{eq:cno2}\\
 3\, {{}^4}\mathrm{He} & \rightarrow & {{}^{12}}\mathrm{C} \label{eq:aaa}\\
 {{}^{12}}\mathrm{C} + {{}^4}\mathrm{He} & \rightarrow & {{}^{16}}\mathrm{O} + \gamma
\label{eq:alpha1}\\
 {{}^{13}}\mathrm{C} + {{}^4}\mathrm{He} & \rightarrow & {{}^{16}}\mathrm{O} + \mathrm{n} \\
 {{}^{14}}\mathrm{N} + {{}^4}\mathrm{He} & \rightarrow & {{}^{18}}\mathrm{O} + \gamma
\label{eq:alpha3}\\
 2\, {{}^1}\mathrm{H} + e^{-} & \rightarrow & {{}^2}\mathrm{H}  + \nu \\
 {{}^{3}}\mathrm{He} + {{}^1}\mathrm{H} & \rightarrow & {{}^4}\mathrm{He} + e^{+} + \nu
\label{eq:last}
\end{eqnarray}
The first five Equations~(\ref{eq:pp}-\ref{eq:be7p})
contain the three alternative
$pp$ branches ($pp1$,$pp2$,$pp3$) all of which start with
${{}^3}\mathrm{He}$. Equations~(\ref{eq:cno1})-(\ref{eq:cno1l})
represent the primary CNO cycle, (\ref{eq:cno2}) the secondary
cycle. The reaction of helium burning is given
in Equation~(\ref{eq:aaa}), followed by the dominant
$\alpha$ capture reactions (\ref{eq:alpha1})-(\ref{eq:alpha3}).
The last two reactions are only important for the neutrino problem and
can be neglected for the energy generation.
\cws{As is implicitly shown in the nuclear
reactions (\ref{eq:pp}-\ref{eq:last}), all
$\beta$-decay reactions
are treated in the instantaneous approximation.}
In addition, four branching
ratios are defined \citep{1988RvMP...60..297B}:
the fraction of ${{}^{7}}\mathrm{Be}$ that is burned by
electron capture (\ref{eq:be7e}),
the fraction of ${{}^{7}}\mathrm{Be}$
that is burned by proton capture (\ref{eq:be7p}),
the fraction of ${{}^{14}}\mathrm{N}$ that is burned via
${{}^{14}}\mathrm{N}(\mathrm{p},\alpha){{}^{12}}\mathrm{C}$ and
the fraction of ${{}^{15}}\mathrm{N}$ that is burned via
${{}^{15}}\mathrm{N}(\mathrm{p},\gamma){{}^{16}}\mathrm{O}$.

The energy generation is calculated by multiplying the rates
by the \cws{$Q$}-values which are taken from
\citet[Table 21]{1988RvMP...60..297B}.
The standard reaction rates implemented are identical to the rates
published in \citet{1989neas.book.....B}.

\subsection{NACRE}\label{nacre}
YREC provides the option of using the
NACRE\footnote{Nuclear Astrophysics Compilation of REaction Rates}
reaction rates \citep{1999NuPhA.656....3A}.
In its present version, the $Q$-values from each
reaction are \emph{not} changed and are thus not identical to the
values published on the NACRE
database\cws{\footnote{\mbox{\cws{http://pntpm.ulb.ac.be/Nacre}}}}.
All relevant reaction rates that are
provided by NACRE are included,
i.e.\ those corresponding to
Equations~(\ref{eq:pp}-\ref{eq:he3he4}; \ref{eq:be7p}-\ref{eq:aaa}).

The NACRE library lists the rate data in tabulated form
and also provides fit-formulas, the latter of which are implemented. The
fit-formulas are accurate by 3\% - 25\% compared to the
tabulated data, with typical deviations of 10\% - 15\%.

Our tests for a standard solar model have shown that a
calibrated standard model is not affected by the NACRE
reaction rates. The largest differences are found in
the neutrino flux of ${^{8}}$B which differs by about
9\%. This difference is comfortably within other theoretical
uncertainties \citep{2004ApJ...614..464B}.
\subsection{Light elements}\label{light}
A switch permits keeping track of nuclear burning of the light elements    
\cws{$^2$H, $^6$Li, $^7$Li and $^9$Be} at the base of the convection zone 
in models of sun-like stars \citep{1990PhDT.........3D}.

\subsection{Neutrino losses}\label{neut}
Neutrino loss rates are taken from the monograph by \citet{1989neas.book.....B}, 
updated by subsequent private communications from the author.
For advanced stages of stellar evolution, the neutrino  
rates from photo, pair and plasma sources from 
\citet{1989ApJ...339..354I} are included. 

\section{Running YREC}\label{runyrec}
YREC can automatically calculate calibrated solar and stellar models. The
user provides a complete set of constraints along with allowable parameter
variations and YREC will search within the chosen parameter space for a
solution. This is especially convenient since the mixing length parameter
and in some cases the helium abundance used to compute stellar models
must first be established from calibrated solar models. The calibrated values
are sensitive to the choice of opacity tables, the equation of state
formulation, the inclusion of diffusion, and the choice of model atmosphere.
\subsection{Calibrated solar models}\label{calsun}
To produce a calibrated solar model the user inputs the age of the Sun and
its primordial composition, i.e., mass fraction mixture of hydrogen, helium,
and metals on the zero age main-sequence. In addition the user specifies
the tolerances for the luminosity and radius. YREC will then vary the
initial value for the helium abundance and mixing length parameter until
it has produced a model at the age of the Sun that has the observed radius,
$6.958\times10^{10}$ cm, and observed luminosity, $3.8515\times10^{33}$ 
$\mathrm{erg/s}$ \citep{1992ApJ...394..313E}  
within the specified tolerances. When including
the effects of metal and helium diffusion, the user has the option of
inputting the $Z/X$ at the surface and its allowed tolerance. In this case,
YREC will adjust the initial helium abundance, metal abundance, and mixing
length until a model at the age of the Sun is produced that matches the
Sun's luminosity, radius, and surface $Z/X$ within the specified tolerances.
With 64-bit floating point numbers, YREC can compute a tuned solar model
with tolerances of 1 part in $10^{6}$ for radius and luminosity and 1 part
in $10^{4}$ in $Z/X$ after about 10 to 12 iterations.

The actual procedure begins by computing one reference run, one run with
slightly changed helium abundance, followed by one run with slightly
changed mixing length parameter, and then one run, if chosen, with
slightly changed metal abundance. The luminosity, radius, and, if chosen,
surface $Z/X$, of the final models are used to compute the derivative
matrix of luminosity, radius and surface $Z/X$ with respect to helium
abundance, mixing length parameter, and \cws{$Z$}.
The first order corrections to
each parameter are determined from the derivative matrix and a new model
is computed. The process is iterated until the model falls within the
specified tolerances.
\subsection{Calibrated stellar models}\label{calstar}
The process is slightly different for stars because normally the age of
a star is unknown. Only the luminosity and surface temperature are
used to constrain the model. In the case of stars, YREC adjusts the
mass and either the mixing length parameter or the helium abundance in
an attempt to produce a stellar evolutionary track that passes though
the tolerance 
box in the theoretical
HR-diagram.

To produce a stellar model the user inputs the metal abundance, and
either the helium abundance or the mixing length parameter. In addition
the user specifies the luminosity, effective temperature, and their
corresponding tolerances. The user has the option of allowing the code
to adjust either the mixing length parameter or the helium abundance.
The code generates tracks varying the mass and the chosen parameter
using a derivative matrix to produce a model that passes through the
specified location in the HR-diagram. Once the optimum parameters are
determined, the code computes the track a second time but stops the
evolution when the model hits the specified location in the HR-diagram.
The tuned model is constrained in mass and age.
\subsection{Pulsation models}\label{pulse}
The pulsation output files in YREC are tailored for the JIG
non-adiabatic oscillation code of Guenther \citep{1994ApJ...422..400G}.
These files can be saved for specified models in an 
evolutionary sequence (say for
a calibrated solar model, or for a calibrated stellar model), or any model for
specified ages along the evolutionary track.

One of the first things stellar modelers realized when using their
solar models for pulsation analysis is that the optimum distribution of
shells within the model for structure and evolutionary calculations is
different from the optimum distribution of shells for pulsation analysis.
Whereas evolutionary models need to resolve well the nuclear burning
regions, pulsation models need to resolve the surface layers
(for acoustic modes). For example, for evolutionary models great care
is needed to fully resolve the thinning hydrogen burning shell
(0.001 $M_{\odot}$ to 0.0001 $M_{\odot})$ 
as the models evolve up the giant branch. For pulsation models it is
the low density regions, where the sound waves have the largest amplitudes,
that need to be well resolved. Therefore, in order to produce viable
models for pulsation analysis, the user increases the resolution of
shells in the envelope, atmosphere, and the region below the base of
the convection zone. Ultimately, in order to achieve frequency accuracies
of the order of 1 part in $10^{4}$ using a first order numerical pulsation
program one needs approximately 600 shells in the interior, 600 shells
in the envelope defined as the outer $1-5\%$ of the mass, and 600 shells
in the atmosphere.
To maximize self consistency all thermodynamic variables and their
derivatives are obtained directly from the structure model.

Related to shell resolution is the distribution of shells near the core.
Stellar evolutionary codes do not locate a shell at the center owing
to divide by zero complications but set the innermost shell a small
distance away from the center. In order to do accurate pulsation
analysis of \cws{$g$-modes} or to study the \cws{$p$-mode} small spacing parameter,
both of which are sensitive to the structure of the deep interior,
it is necessary to extend the innermost shell closer to the center
than normally required by stellar evolutionary calculations:
compare $1.0\times10^{-3}$ radius fraction for stellar evolution to
$2.0\times10^{-7}$ for stellar pulsation.
A stellar model output for pulsation runs from the ``central-most''
interior shell to the top of atmosphere computation near 
optical depth $\tau=10^{-10}$.
\subsection{Model grids}\label{grids}
A useful feature of YREC is its ability to carry out extensive model
calculations without user input. It is possible to generate in a single
run, tens of thousands of evolutionary tracks, corresponding to tens
of millions of models, covering a wide range of masses, compositions,
mixing length parameters, with each track tuned to their own
particular numerical and physical variables. This has enabled
\citet{2004ApJ...600..419G} to
compute dense grids of stellar models for pulsation analysis throughout
the HR-diagram.  For other grids of evolutionary sequences, see 
Sect.~\ref{other}.
\subsection{Backwards compatibility}\label{backw}
All new physics (e.g., opacities, nuclear reaction rates) have been
implemented along with existing physics so that the user can, at any time,
run YREC using older physics. 

\section{Convection}\label{conv}
By default the local Schwarzschild criterion is implemented
in order to determine if a mass shell is labeled as \emph{convective}
or \emph{radiative}. The Ledoux stability criterion can also
be used when a specific parameter choice is made in
the local limit of the non-local convection treatment described below.
The abundance of chemical species in convective cores is
treated under the assumption of instantaneous mixing.

\section{Core Overshoot}\label{ov}
Since the eddy velocity at convective boundaries is non-zero,
convective motions will penetrate into the radiative region. Two
different forms of penetration are commonly distinguished: (a)
inefficient penetration that does not alter the temperature gradient,
termed ``overmixing'' here, and (b) subadiabatic penetration
\citep{1991A&A...252..179Z}, where the convective heat transport is
efficient enough to establish a nearly adiabatic temperature gradient.

YREC offers a number of different options for treating overshoot
(OS). All OS options have in common that mixing of chemical species in
the OS region is instantaneous and all chemical species are
homogenized within the extended zone. Due to the small characteristic
time scale of convection in comparison to the thermal and nuclear
timescales during the major burning stages this assumption holds to
high accuracy.

Among the OS options, two different approaches are
distinguished: (a) a parametric treatment where the OS extent is a
multiple of the pressure scale height taken at the formal
Schwarzschild boundary and (b) a physically motivated treatment where the
OS extent is calculated from a non-local convection theory
originally developed by \citet{1986A&A...160..116K} and
later extended by \citet{1998A&A...340..419W}.  In the
latter case, the temperature gradient is calculated directly from the
additional convection equation which is solved in addition to the
canonical stellar structure equations at every time-step.
\subsection{Parametric Treatment}\label{param}
The boundary of the OS zone is determined by adding a fraction
$\alpha_{\mbox{\scriptsize{\cws{OM}}}}$
of the pressure scale height to the boundary at
the radius $r_{\mbox{\scriptsize{\cws{S}}}}$,
determined by the Schwarzschild criterion:
\begin{equation}
r_{\mbox{\scriptsize{\cws{new}}}} = r_{\mbox{\scriptsize{\cws{S}}}} +
\alpha_{\mbox{\scriptsize{\cws{OM}}}}\, H_P(r_{\mbox{\scriptsize{\cws{S}}}})
\end{equation}
where the pressure scale height $H_P(r_{\mbox{\scriptsize{\cws{S}}}})$
is taken at the Schwarz\-schild boundary. The
temperature stratification in the OS zone is determined by
the two options described above, either (a) the temperature gradient is
not altered (overmixing) or (b) the temperature gradient is set to
the adiabatic temperature gradient. For a fixed
$\alpha_{\mbox{\scriptsize{\cws{OM}}}}$
the latter option produces larger convective cores
\citep{1991A&A...252..179Z}.
\subsection{Non-local Convection}\label{nonlocal}
As an alternative to the purely parametric treatment of OS, the
one-dimensional convection theory developed by
\citet{1986A&A...160..116K} is
implemented \citep{2005ApJ...629.1075S}.
In the framework of anelastic and diffusion-type
approximations of the unknown correlation functions, Kuhfu{\ss} derives
one equation for the turbulent kinetic energy from spherical averages
of the first-order perturbed Navier-Stokes equations. The solution of
this equation provides the extent of the convective core region and
includes the effects of OS naturally, since the velocity of
convective motions is zero where the turbulent kinetic energy
vanishes. In addition, this equation also provides the temperature
gradient in the OS region.
\subsubsection{Implemented Equations}\label{impeq}
The new equation for the turbulent kinetic energy $\wturb$ that is solved
in YREC is given by:
\begin{eqnarray}
  \lderivt{\wturb} & = & \frac{\nablad}{\rho \Hp}\, \jw
  - \frac{c_D}{\Lambda}\, \wturb^{3/2}
  - \gradm{}\left(4 \pi r^2 j_\mathrm{t} \right) \komma
\label{eq:wturbeqnsimp} \\
\jt & = & - 4 \pi r^2 \rho^2 \alpht\, \Lambda\, \wturb^{1/2}\, \gradm{\wturb}
\end{eqnarray}
where $\mbox{D}/\mbox{D}t$ is the Lagrangian time derivative,
$\rho$ density, $r$ stellar radius,
$m$ the Lagrangian mass coordinate. $\Lambda$, defined as
$1/\Lambda=1/(\alphMLT \Hp)+1/(\betr r)$ is the
geometrically limited mixing length scale with
$\nablad$ being the adiabatic gradient. Note that the
limiting of the pressure scale height in the central
part influences the total core size within the framework of
non-local convection theories.

A linear implicit extrapolation method is used in order to calculate the
stationary solution of Equation~(\ref{eq:wturbeqnsimp}), i.e. 
$\mbox{D}\wturb/\mbox{D}t \equiv 0$. The solution yields the
turbulent kinetic energy $\wturb$ at every mass shell. We define shells to
be convective, if:
\begin{equation}
x_{\wturb} < 0.1
\end{equation}
where
\begin{equation}
x_{\wturb} = \frac{1}{1 + F\, \wturb^{1/2}} \komma\qquad
F = \frac{3 \alphs\, \kappa\, \rho^2\, \Lambda \Cp}{16 \sigma T^3}
\end{equation}
with the usual notation for the opacity $\kappa$,
temperature $T$, specific heat at constant pressure $\Cp$ and
the Stefan-Boltzmann constant $\sigma$.
The boundaries are sharply defined by an extreme falloff
of $\wturb$ which is encountered in interior solutions of
Equation~(\ref{eq:wturbeqnsimp}). 
Finally, the temperature gradient can be calculated from:
\begin{equation}
\nabla = \nablad + x_{\wturb}\, (\nablrad-\nablad) + (1-x_{\wturb})
         \, \left( G \gradn{\wturb} + H \right)
\end{equation}
with
\begin{eqnarray}
G &=& \frac{\alpht }{\alphs}\, \frac{\Hp}{\Cp\, T} \left(\frac{4 \pi r^2 \rho}{\delM}\right) \komma \\
H &=& \frac{\varphi}{\delta} \nablmu
\end{eqnarray}
where $\delM$ is the mass enclosed in one shell and
$\nablmu= d\ln \mu/d\ln P$. In the
case of an ideal gas with radiation pressure
the dimensionless parameters $\delta$ and $\varphi$ take on the
values $\delta = (4-3\beta)/\beta$, $\beta = \Pgas/P$ and
$\varphi=1$ respectively.
In convective core regions, the temperature gradient remains
very close to the adiabatic one whenever $x_{\wturb} < 0.1$.
A more detailed discussion of the 
implemented equations and the numerical techniques employed
can be found in \citet{2005ApJ...629.1075S}.
\subsubsection{Non-local parameters}\label{nlparam}
\begin{table}[t]
\caption{Parameters of non-local convection theory}
\centering
\label{tab:para}
\begin{tabular}{lll}
\hline\noalign{\smallskip}
parameter & canonical value & description  \\[3pt]
\tableheadseprule\noalign{\smallskip}
$\alphMLT$ & $1.5$    & mixing length\\
$\alphs$   & $0.408$  & turbulent driving\\
$c_D$      & $2.177$  & dissipation efficiency\\
$\alpht$   & $0.610\,\alphs$    & overshoot distance\\
$\betr$    & $1.0$    & geometric mixing length\\
\noalign{\smallskip}\hline
\end{tabular}
\end{table}
The implemented non-local convection theory contains five
parameters (Table~\ref{tab:para}).
These parameters must be calibrated, preferably
on a well selected set of open clusters, or on selected
asteroseismic target stars like Procyon~A \citep{2005ApJ...629.1075S}.
Two of the canonical values given in Table~\ref{tab:para},
i.e. $\alphs$ and $\alpht$ can be derived by matching the Kuhfu{\ss}
result to \emph{mixing length theory}
(MLT) in the local limit ($\alpht = 0$).

The mixing-length parameter plays a minor role in the core regions,
where superadiabaticity of temperature gradient is tiny. The
parameter that controls the OS zone is given by
$\alpht$ and is thus the most important one to calibrate.
Kuhfu{\ss} derives $\alpht = 0.610\,\alphs$ from theoretical
arguments.

In the strictly local limit ($\alpht = 0$) the Kuhfu{\ss} treatment is
equivalent to the MLT equations when based on the Ledoux
stability criterion.

\section{Turbulence}\label{turb}
A method to incorporate the effects of turbulence into the 
outer layers of one-dimensional (1D) 
stellar models has been implemented in YREC \citep{2002ApJ...567.1192L}.
The method requires a detailed three-dimensional 
hydrodynamical simulation (3D RHD) of the atmosphere and highly
superadiabatic layer of stars
\citep{2003MNRAS.340..923R}.

The basic idea is to extract from the velocity field of the 3D simulation
three important quantities: the turbulent pressure, the turbulent kinetic
energy and the anisotropy of the flow.  
Implementation into a 1D stellar model thus requires 
two additional parameters, i.e. $\chi$, the specific turbulent
kinetic energy, and $\gamma$, which reflects the flow anisotropy.
These parameters, which modify
the hydrostatic equilibrium equation and the internal
energy equation, must be introduced
in a thermodynamically self-consistent way.  As a result, they also
change the
adiabatic and convective temperature gradients, as well as the energy
conservation equation.

The next section (Sect.~\ref{tvel}) describes the calculation of $\chi$ and
$\gamma$ from the velocity field in the 3D simulation.  The introduction of
the parameters $\chi$ and $\gamma$
into the stellar structure equations and YREC
is summarized in Sect.~\ref{s3}, Sect.~\ref{app1}
and Sect.~\ref{app2}.  The effects on $p$-mode frequencies in a solar model
are illustrated in Sect.~\ref{fshift}.

\subsection{Turbulent velocities}\label{tvel}
The physics (thermodynamics, the equation of state, and opacities) in the 3D simulation is
the same as in the 1D stellar models. These simulations follow closely
the approach described by \citet{1998ApJ...496L.121K}, and are
described in more detail in the papers of
\citet{2003MNRAS.340..923R,2004MNRAS.347.1208R}. The full hydrodynamical
equations were solved in a thin
subsection of the stellar model, i.e. a 3D box located in the vicinity of the
photosphere. For the radiative transport, the diffusion approximation
was used in the deep
region ($\tau > 10^3$) of the simulation, while the 3D Eddington approximation
was used \citep{1966PASJ...18...85U} in the region above. 
After the simulation
had reached a steady state, statistical integrations were performed for each
simulation for over 2500 seconds in the case of the solar surface
convection.

For the derivation of $\chi$ and $\gamma$, \citet{2002ApJ...567.1192L}
use a self-consistent approach introduced by
\citet{1995ApJS..101..357L} to include magnetic
fields in calculating the convective temperature gradient
within the MLT framework, and used successfully by \citet{1996ApJ...456L.127L} to
explain the variation of solar $p$-modes with the solar cycle.

Turbulence can be measured by the turbulent Mach number ${\cal M}=v''/v_s$,
where $v''$ is the turbulent velocity, and $v_s$ is the sound speed.
The MLT is valid
when ${\cal M}$ is sufficiently small. In the outer 
layers of a star like the sun ${\cal M}$
can be of order unity \citep{1968QB801.C65}, but in the 
deep convection region
${\cal M}$ is almost zero. The turbulent velocity is defined by the velocity
variance:
\begin{equation}
  v''_i = (\overline{{v_i}^2}-\overline{{v_i}}^2)^{1/2}, \label{eq:rms}
\end{equation}
where the overbar denotes a combined horizontal and temporal average, and $v_i$
is the total velocity.

Using ${\cal M}$, we can define the
turbulent kinetic energy per unit mass $\chi$ as
\begin{equation}
  \chi = \cws{\frac{1}{2}}\, {\cal M}^2 v_s^2.
\end{equation}
The turbulent contribution to the entropy is
\begin{equation}
  S_{\mbox{\scriptsize{turb}}} = \chi/T,
\end{equation}
where $T$ is the gas temperature.

Turbulence in the stratified layers of a stellar convection zone is not isotropic.
We define the parameter $\gamma$ to
reflect the anisotropy of turbulence,
\begin{equation}
  P_{\mbox{\scriptsize{turb}}}= (\gamma-1)\rho\chi, \label{eq:pturb}
\end{equation}
where $\rho\chi$ is the turbulent kinetic energy density.
Since $\cws{P_{\mbox{\scriptsize{turb}}}} = \rho {v''_z}^2$,
$\gamma$ can be related to the turbulent velocity as follows:
\begin{equation}
  \gamma = 1 + 2(v''_z/v'')^2.
\end{equation}
$\gamma = 5/3$ when turbulence is isotropic ($v''_z=v''_x=v''_y$); $\gamma=3$ or
$\gamma=1$ when turbulence is completely anisotropic ($v''_z=v''$ or $v''_z=0$,
respectively). The physical meaning of $\gamma$ is the 
specific heat ratio due to
turbulence. 
This affects the distribution of the radial turbulent pressure
which is then scaled with
the gas pressure,
$P_{\mbox{\scriptsize{gas}}}$. 
The total pressure is defined as
\begin{equation}
P_T=P_{\mbox{\scriptsize{gas}}}+P_{\mbox{\scriptsize{rad}}}
+P_{\mbox{\scriptsize{turb}}}.
\end{equation}

\subsection{Convective temperature gradients with the turbulent
velocities}\label{s3}
\myfigure{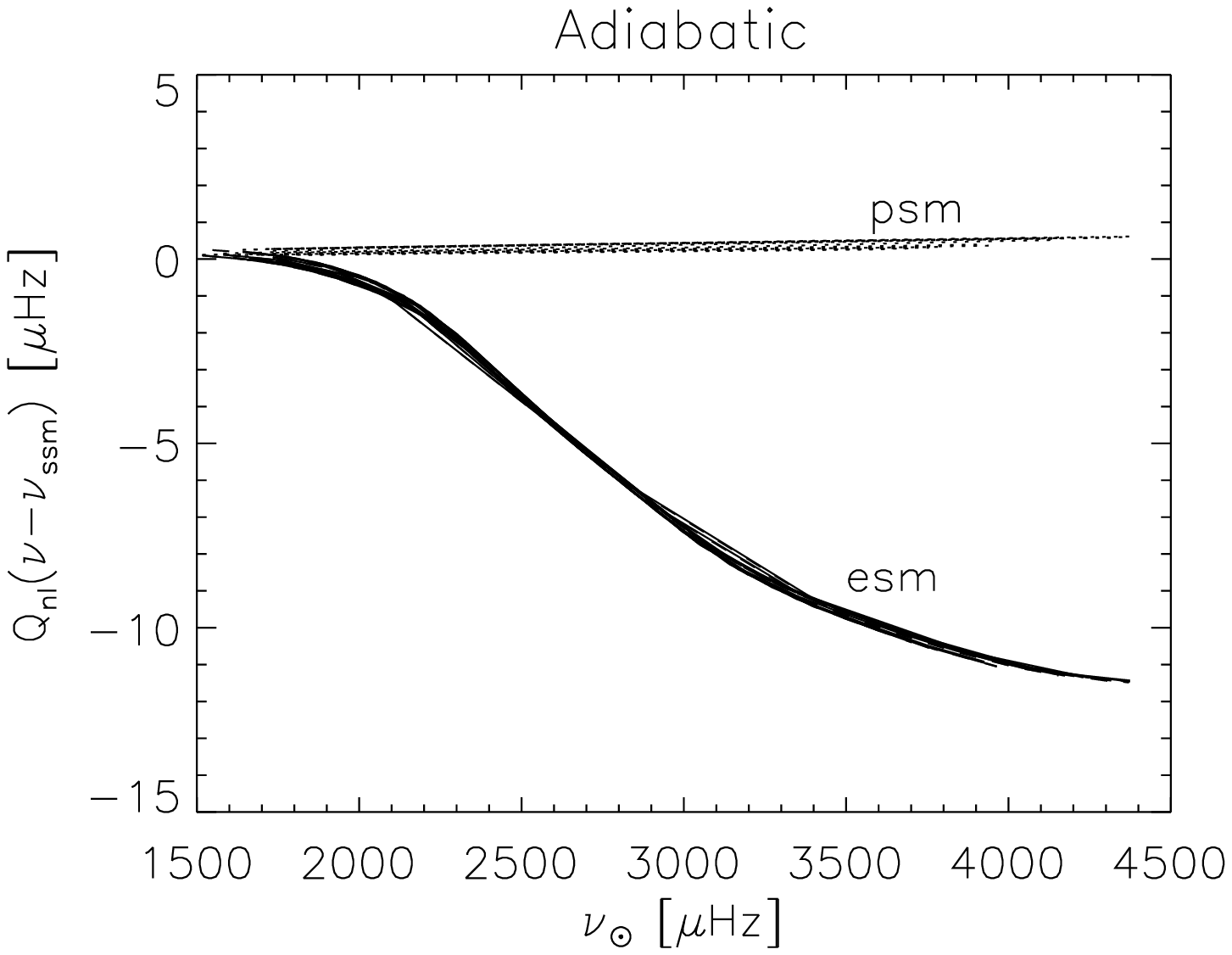}{
\cws{$P$-mode} frequency difference diagrams. Turbulent model minus 
standard model (ssm), for
the turbulent pressure solar model (psm), and the solar model 
with the turbulent pressure
and turbulent kinetic energy (esm). The difference between psm and ssm 
is of the order of 
\cws{$1 \mu\,$Hz}, while at high frequencies esm and ssm differ by 
more than \cws{$10$ $\mu\,$Hz}.  
Plotted are the \cws{$l = 0$, $1$, $2$, $3$, $4$,
$10$, $20$, $\ldots$, $100$ $p$-modes}.
\label{turbdif}
}
Since the parameters 
$\chi$ and $\gamma$ now appear in the equation of state, 
they must be included as independent
variables in evaluating the density derivative.  We have therefore:
\begin{equation}
 d\rho/\rho = \mu d P_T/P_T - \mu' dT/T - \nu d\chi/\chi - \nu' d\gamma/\gamma,
\end{equation}
where
\[
\begin{array}{ll}
\mu = \left(\p{\ln\rho}{\ln P_T}\right)_{T,\chi,\gamma} &
\mu' = -\left(\p{\ln\rho}{\ln T}\right)_{P_T, \chi, \gamma}  \\
\nu = -\left(\p{\ln\rho}{\ln \chi}\right)_{P_T, T, \gamma}  &
\nu' = -\left(\p{\ln\rho}{\ln \gamma}\right)_{P_T, T, \chi}   \\
\end{array}
\]

As a result, the stability criterion against convection is modified.  For 
similar reasons, both
the convective and adiabatic gradients
 are also modified by turbulence.

\subsection{Solar model with turbulent pressure alone}\label{app1}

The simplest way to take into account turbulence in solar modeling is to
include turbulent pressure (or Reynolds stress) alone.
In this case, only the hydrostatic equilibrium equation
needs to be modified as follows:
\begin{equation}
    \p{P}{M_r} = - \frac{GM_r}{4\pi r^4}(1+\beta),
\end{equation}
where $P=P_{\mbox{\scriptsize{gas}}}+P_{\mbox{\scriptsize{rad}}}$, and
\begin{equation}
  \beta = \left(\frac{2P_{\mbox{\scriptsize{turb}}}}{\rho g
r}-\p{P_{\mbox{\scriptsize{turb}}}}{P}
    \right) \left(1+\p{P_{\mbox{\scriptsize{turb}}}}{P} \right)^{-1}.
\label{eq:hydrostatic}
\end{equation}
Here $2P_{\mbox{\scriptsize{turb}}}/(\rho g r)$ originates from the spherical
coordinate system adopted, representing a kind of geometric effect. The
equations that govern the envelope integrations also need to be changed accordingly.

One can construct a calibrated
nonstandard model in the same way as one obtains the standard solar model, assuming
that $P_{\mbox{\scriptsize{turb}}}$, set equal to its value for the present sun,
does not change from the ZAMS to the present age of the sun.
The $p$-mode oscillation spectrum of this calibrated solar model (psm) is discussed in 
Sect.~\ref{fshift}.
\myfigure{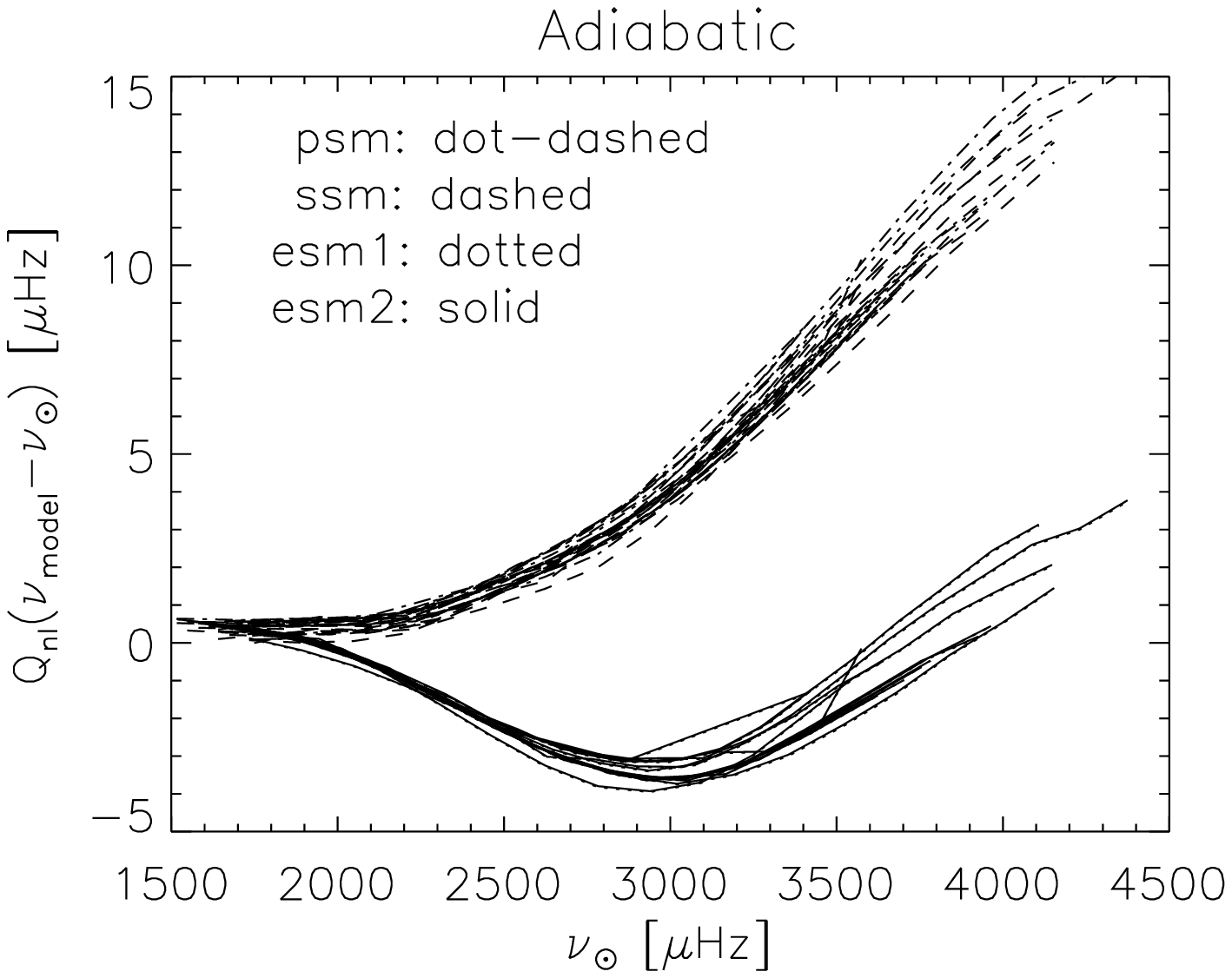}{
\cws{$P$-mode} frequency difference diagrams, observation minus model, scaled by the
mode mass $Q_{nl}$, for the standard solar model (ssm), the turbulent pressure
solar model (psm), a solar model with fixed turbulent pressure and kinetic
energy (esm1), and a solar model with evolutionary turbulent pressure and
kinetic energy (esm2, almost overlaps with esm1). Plotted
are the \cws{$l = 0$,
$1$, $2$, $3$, $4$, $10$, $20$, $\ldots$, $100$ $p$-modes}.
\label{pmode}
}

\subsection{Solar model with $\chi$ and $\gamma$ as independent 
variables}\label{app2}

The form of the continuity equation and of the equation of 
transport of energy by radiation
are not affected by turbulence. The hydrostatic equation includes a
Reynolds stress term due to turbulence
\begin{equation}
    \frac{\partial P}{\partial r} = - \frac{GM_r}{r^2}\rho -
\frac{1}{r^2}\od{}{r}(r^2\rho v_rv_r),
\end{equation}
where $P=P_{\mbox{\scriptsize{gas}}}+P_{\mbox{\scriptsize{rad}}}$. Since the
last term can be rewritten as
$\partial P_{\mbox{\scriptsize{turb}}}/\partial r+2(\gamma-1)\chi/r$, this
equation becomes
\begin{equation}
  \p{P_T}{M_r} = - \frac{GM_r}{4\pi r^4} - \frac{2(\gamma-1)\chi}{4\pi r^3}.
  \label{eq:hstate}
\end{equation}
The last term on the right hand side of Eq.~(\ref{eq:hstate}) also embodies
the same spheric geometric effect as $2P_{\mbox{\scriptsize{turb}}}/(\rho g r)$
in Eq.~(\ref{eq:hydrostatic}).

The energy conservation equation is also modified by
turbulence because the first law of thermodynamics must now include the turbulent
kinetic energy.  We have then:
\begin{equation}
  \p{L_r}{M_r} = \epsilon - T\od{S_T}{t}, \label{eq:energy}
\end{equation}
where
\begin{equation}
  TdS_T = c_{\rm p}dT - \frac{\mu'}{\rho} dP_T + \left(1+\frac{P_T\mu'
\nu}{\rho\mu\chi}\right)d\chi +
                       \frac{P_T\mu'\nu'}{\rho\mu\gamma}d\gamma.
\label{eq:firstlaw}
\end{equation}

The equation of energy transport by convection,
\begin{equation}
  \p{T}{M_r} = - \frac{T}{P_T}\frac{GM_r}{4\pi r^4}
\nabla_{\mbox{\scriptsize{conv}}},
\label{eq:convection}
\end{equation}
does not change in form, but the convective temperature gradient, discussed in
a previous section, is different from that without turbulence. The equations
that govern envelope integrations also need to be changed accordingly.
The oscillation properties of the calibrated solar model 
constructed under this assumption (esm) are discussed in the next
section.

\subsection{Frequency corrections to solar $p$-modes}\label{fshift}
Implementing the effects of turbulence in the outer layers 
of the stellar model modifies the calculated $p$-mode frequencies at 
high frequencies.  The magnitude of the frequency correction is illustrated 
in Fig.~\ref{turbdif} and Fig.~\ref{pmode} for the case of a 
solar model, taken from the work of \citet{2002ApJ...567.1192L}.
In the \citet{2002ApJ...567.1192L} paper, the $p$-mode frequencies for two  
calibrated solar models that include the effects of turbulence 
are compared to the standard solar model (ssm) $p$-mode frequencies.
The psm model is obtained by including turbulent pressure
alone in the solar modeling, while the esm models are obtained by 
introducing the turbulent
variables $\chi$ and $\gamma$ which include both turbulent pressure 
and kinetic energy. Contrary to a frequently made assertion, the inclusion 
of turbulent pressure in the pressure term has only a small 
effect on the calculated $p$-mode 
frequencies.  On the other hand, the inclusion of turbulent kinetic 
energy is significant.  This is illustrated in
Fig.~\ref{turbdif} which shows that the frequency
differences caused by turbulent kinetic energy are much larger in size 
than those caused
by turbulent pressure alone. Fig.~\ref{pmode} indicates that the frequency
changes caused by turbulent kinetic energy make the computed model frequencies
match the solar data better than the ssm model. This result is 
consistent with the work of \citet{1999A&A...351..689R} who ``patched'' a 
modified 3D RHD simulation by \citet{1989ApJ...342L..95S} onto a 1D solar model 
(see their Figures~1 and 6).

\subsection{SAL peak shift}\label{salpeak}
While it is always preferable to extract the $\gamma$-$\chi$ data from
a 3D RHD simulation that corresponds to exactly the same atmospheric
conditions ($\log~g$, \cws{$\log~\Teff$}) as in the
\cws{1D} model, it
\cws{is of} interest to estimate the turbulence effects in stellar
models where the 3D RHD simulation is \cws{not available}.

In such situations, the $\gamma$-$\chi$ data cannot be used
directly, instead the data must be shifted in order to be applied at
the correct depth of the model \citep{2006ApJ...636.1078S}. This 
shifting is motivated by an
expected characteristic found in all 3D RHD atmosphere
simulations: namely that
the SAL peak closely coincides with the turbulent pressure peak.

\subsection{Calibration}\label{calib}
The presented method for including turbulence does not remove MLT and
therefore
the uncertainties inherent in the mixing length parameter remain. In
order to make quantitative predictions, both mass and age of the star
must be known to high precision in addition to the luminosity $L$ and
effective temperature \cws{$\Teff$}. Instead of the latter an interferometric
radius measurement is preferable, since the measurement is
usually more precise.

In the case of the Sun, the age is known to high precision and the mixing
length parameter and hydrogen mass fraction can be calibrated to the
known solar luminosity and radius.  As demonstrated in
\citet{2005ApJ...635..547G,2006ApJ...636.1078S}, asteroseismic data can
be instrumental in other stars, since low order $p$-mode frequencies
anchor the interior model effectively in age and mass. When calibrated
to the same luminosity and effective temperature, a differential
assessment of turbulence effects can be derived.

Another example, in which a proper calibration is possible with more
asteroseismic data, is the detached binary system $\alpha$ Centauri: The
masses of both components are known to high precision, the radius of
the $A$ component is measured with interferometric techniques
\citep{2003A&A...404.1087K} and the luminosity is also well determined
through parallax measurements. With the help of future asteroseismic
data of the low order $p$-mode frequency spectrum, 
the stellar age of $\alpha$ Centauri can be effectively determined.
Under such circumstances the methods described to include
turbulence in YREC are fully applicable.

\section{Seismic diagnostics}\label{dia}
Stellar models constructed with YREC have been used to develop seismic 
diagnostics to explore internal structural properties of stars that could not be 
observed by any other means. 

\citet{2004MNRAS.350..277B} have used low degree acoustic modes to determine the 
\emph{helium abundance} in the envelopes of low-mass main sequence stars 
with precision.  The oscillatory 
signal in the frequencies caused by the depression in $\Gamma_{1}$ in the second 
helium ionization zone is used.  For frequency errors of one part in $10^4$, the 
expected $\sigma_{Y}$ in the estimated \cws{$Y$} ranges from 0.03 for $0.8 M_{\odot}$ stars 
to 0.01 for $1.2 M_{\odot}$ main sequence stars. In 
more evolved stars, this approach  
is feasible if the relative errors in the frequencies are less than $10^{-4}$.   

\citet{2006MNRAS.372..949M} have explored asteroseismic 
diagnostics of \emph{convective core mass} 
using small frequency separations of low-degree $p$-modes.  Small separations 
can also be combined to derive convective \emph{core overshoot} 
diagnostics. It was shown that in stars 
with convective cores, the mass of the convective core can be estimated to within 
$5\%$ if the total mass is known, although systematic errors in the total mass 
could introduce errors of up to $20\%$. The evolutionary 
stage of the star, determined 
in terms of the central hydrogen content is much less sensitive to the mass 
estimate.   

\section{Solar neutrinos and helioseismology}\label{solar}
Different versions of YREC have been used to study the structure of the 
solar interior, the solar neutrino problem and helioseismology 
\citep{1997ApJ...484..937G}
and references therein. 
In particular, the important series 
of papers by 
\citet{2004ApJ...614..464B,2005ApJ...618.1049B} 
on solar neutrinos, 
helioseismology and solar abundances also made use of a dedicated version  
of YREC.
 
\section{Other YREC applications}\label{other}
A variety of applications to stellar structure theory and evolution 
have been carried out using YREC.  In addition of the work on stellar rotation 
mentioned in the introduction (see also \citet{1995ApJ...441..865C} and 
\citet{2003ApJ...586..464B}), one notes the
pioneer work on 
stellar collisions and mergers \citep{1997ApJ...487..290S}.

Important research in stellar population studies and population synthesis 
continues to be 
carried out with the Yonsei-Yale isochrones 
(YY isochrones) 
\citep{2001ApJS..136..417Y}.
Frequently quoted research
on helium burning phases of evolution (hor\-i\-zon\-tal-branch) has also been 
carried out with 
YREC \citep{1994ApJ...423..248L,1997ApJ...482..677Y}\cws{.}
 
\begin{acknowledgements}\label{ack}

A large number of people have contributed to YREC, in ways large and small. 
In addition to the authors of this write-up, researchers currently at Yale 
who are doing YREC related research
include S. Basu and students, F.J. Robinson, and S. Sofia.  
There is an on-going collaboration with researchers at other institutions:
co-author D.B. Guenther and his students at Saint Mary's University,
Y.-C. Kim, Y.-W. Lee and S.K. Yi and their students at Yonsei University,
and co-author C.W. Straka at CAUP (Porto).

Thanks are due to M. Pinsonneault now at Ohio State University, who 
is the main architect of the 
original YREC created from previous Henyey codes at Yale. He and his students 
have developed their own version of YREC.  So have A. Sills at McMaster University 
and B. Chaboyer at Dartmouth College and their students and collaborators.
S. Barnes at Lowell Observatory and C. Deliyannis at
Indiana University also use YREC in their research.  J.-H. Woo has worked on 
core overshoot.

Partial support from NASA Astrophysics Theory grants NAG5-8406 and NAG5-13299  
and from HST-GO-10505.03-A is gratefully acknowledged by members of the Yale group. 
DBG acknowledges 
the support of the Natural Sciences and Engineering Council of Canada.
\cws{CWS acknowledges support from the European
Helio- and Asteroseismology Network (HELAS) funded by
the European Union's Sixth Framework Program.}

We would like to thank the anonymous referee for doing a meticulous job
that provided very helpful comments to improve the manuscript.
\end{acknowledgements}
\vspace*{-6mm}
\bibliographystyle{aa}
\bibliography{apj-jour,esta15}   
\end{document}